\documentclass[a4paper,fleqn,usenatbib]{mnras}

\usepackage{newtxtext,newtxmath}

\usepackage[T1]{fontenc}
\usepackage{ae,aecompl}

\usepackage{graphicx}	
\usepackage{amsmath}	

\newcommand{\revI}[1]{#1}
\newcommand{\rchange}[1]{#1}
\newcommand{\cchange}[1]{#1}
\newcommand{\al}{{$^{26}$Al~}}
\newcommand{\alu}{{$^{26}$Al}}
\newcommand{\fe}{{$^{60}$Fe~}}
\newcommand{\feu}{{$^{60}$Fe}}

\newcommand{\mspy}{{M$_\odot$~yr$^{-1}$}}

\newcommand{\kms}{{km~s$^{-1}$}}

\newcommand{\rvb}[1]{{ #1}}

\newcommand{\fext}{.png}

\newcommand{\fdirthreeD}{./}
\newcommand{\fdir}{./}

\title[Tracing metal loss with \al]{Galactic \al traces metal loss through hot chimneys}
\author[M. G. H. Krause et al.]{
Martin G. H. Krause$^{1}$\thanks{E-mail: M.G.H.Krause@herts.ac.uk}, 
Donna Rodgers-Lee$^{2}$,
James E. Dale$^{1}$,
Roland Diehl$^{3}$  \newauthor and
Chiaki Kobayashi$^{1}$\\
$^{1}$Centre for Astrophysics Research, Department of Physics, Astronomy and Mathematics, University of Hertfordshire, 
College Lane, \\ \:\,Hatfield, Hertfordshire AL10 9AB, UK\\
$^{2}$School of Physics, Trinity College Dublin, University of Dublin, 
College Green, Dublin 2, D02 PN40, Ireland\\
$^{3}$Max-Planck-Institut f\"ur extraterrestrische Physik, Giessenbachstrasse 1, Garching D-85748, Germany\\
}

\date{Accepted XXX. Received YYY; in original form ZZZ}

\pubyear{2019}

\begin{document}
\label{firstpage}
\pagerange{\pageref{firstpage}--\pageref{lastpage}}
\maketitle

\begin{abstract}
Radioactive \al is an excellent tracer for metal ejection in the Milky Way, and can provide 
a direct constraint on the modelling of supernova feedback in galaxy evolution.
Gamma-ray observations of the \al decay line have found high velocities and hence require 
a significant fraction of the Galactic \al in the hot component. At the same time, meteoritic data
combined with simulation results suggest that a significant amount of \al makes its way into stars before decay.
We investigated the distribution into hot and cold channels with a simulation of a Milky-Way-like galaxy
with massive-star feedback in superbubbles and with ejecta traced by \alu.
About \rvb{30-}40~per cent of the ejecta 
remain hot, with typical cooling times of the order Gyr. \revI{\al traces the footpoints of a 
chimney-fed outflow that mixes metals turbulently into the halo of the model 
galaxy on a scale of at least  50~kpc.}
The rest diffuses into cold gas \revI{$\lesssim 10^4$~K}, and may therefore be quickly available for star formation.
We discuss the robustness of the result by comparison to a simulation with \rvb{a} different global flow pattern.
The \revI{branching ratio} into hot and cold components is comparable to that of longer term average results
from chemical evolution modelling of galaxies, clusters and the intracluster medium.
\end{abstract}



\section{Introduction}\label{sec:intro}
Observations of galaxy clusters and the intracluster medium
\citep{BW10,RA14,Simionea19}, as well as analysis of metals in stars and the 
interstellar and circumgalactic medium \citep[][and references therein]{MM19}, 
suggest that the majority 
of metals produced in stars probably leaves the galaxies and enriches the 
\cchange{circum-/intergalactic} medium.
The argument is made globally, analysing mass budgets of either
the intracluster medium, or late populations of stars that have been enriched
by many \rvb{preceding} generations of stars over many Gyr. 
\cchange{Chemodynamical cosmological simulations relate the fraction of metals
not quickly recycled in stars to the strength of feedback in the galaxy, and 
predict that this fraction is higher for lower-mass galaxies 
\citep*[e.g.,][]{KSW07,TK15}.}

These extragalactic findings have an important complement in direct 
observations of the ejecta in the Milky Way, where we can trace their 
path \rchange{in gamma rays} with radioactive elements ejected from stars together with the 
stable elements. 
Since chemical elements generally form in stars, radioactive elements 
are a reliable way to identify recent stellar ejecta.
Isotopes with different
decay times trace ejecta from timescales of weeks
\citep[$^{56}$Ni,][]{Diehlea14a} through centuries \citep[$^{44}$Ti,][]{Iyudea94,Renaudea06} and Myr \citep[$^{26}$Al and $^{60}$Fe,][]{Diehlea95,Kniea04,WangWea07,WangWea20}, up to $10^8$~yr \citep[$^{244}$Pu,][]{Wallnea15}.

\rchange{\al ($\tau=1$~Myr) has been observed to be present 
throughout the Milky Way, via the 1.8~MeV radioactive decay line 
\citep{PD96,Plea01,Diehlea06,BJR15}. The appearance of diffuse but clumpy \al emission along
the entire band of the Milky Way has been used to argue that \al is emitted
mainly by massive stars \citep{PD96}. These represent a young stellar population, 
which is forming currently in the disc of the Galaxy, rather than in the bulge. 
\al enters the the interstellar medium with hot gas ejected in 
fast ($\gtrsim 1000$~km~s$^{-1}$) winds
and supernova explosions \citep{PD96,Diehl13}. }
The radioactive decay effectively attaches a clock to the ejected \alu. 
\rchange{High-resolution spectroscopy of the \al gamma-ray line with the spectrometer 
onboard the INTEGRAL satellite has found Doppler shifts and line broadenings. 
Through this, discrimination is obtained of the dominant gas phase that the nuclei reside in at the time of their decay
\citep{Kretschea13}.
\rchange{The high velocities found}
are evidence for a significant part of the Galactic \al production occurring and} remaining 
in the hot phase $\approx 10^6$ years after ejection \citep{Krausea15a}.

\rchange{High abundances of $^{26}$Mg, the decay product of \alu, have been found in early-formed parts of meteorites \citep{MacPhea10,Groopmea15}. This suggests
the inclusion of very recent massive star ejecta in the material from which the solar system formed
\citep[e.g.,][and references therein]{Gounea15,Lugarea18}. \al also may have 
played an important role in heating the protoplanetary disc of the solar nebula, with a major impact on ice melting and evaporation and hence
the abundance of water, crucial for the habitability of rocky planets \citep{Lichtea19}. 
It is debated if the \al abundance in the early solar system was very special or if it is a rather typical case 
\citep[e.g.,][]{Dwarkea17,PZ19}. If it was typical, as recently argued from
3D hydrodynamic simulations \revI{by} \citet{FKT18}, this would 
indicate that also a significant part of massive-star ejecta might be 
incorporated into new stars before \al has decayed, and hence a rapid re-cycling time for this
fraction of the massive-star ejecta.}
 
 \rchange{The study of chemical evolution in galaxies and the intracluster medium thus leads us to a similar picture as the
 \al gamma-ray observations and the data from the early solar system: high fractions of massive star 
 ejecta need on the one hand to be mixed into the dense, star-forming interstellar medium locally
 and on the other hand to remain in hot gas to be able to
 be ejected from the galaxy. Propagation of ejecta
 and mixing of different phases of the interstellar medium can be studied in three-dimensional 
 hydrodynamics simulations \citep[e.g.,][]{Breitschwea16,Krausea18b}. Separate whole-galaxy simulations
 have recently demonstrated that both,  the Milky Way's \al gamma-ray
 emission \citep{RodgLea19} and the amount of \al in dense, star-forming gas 
 can be reproduced,
 with the abundance level of the solar system as a likely outcome \citep{FKT18}. 
 Here, we \revI{now demonstrate explicitly} with representative simulation\revI{s} that comparable fractions of 
 massive-star ejecta traced by \al
 are indeed found in cold and hot gas, respectively,
 \revI{and that \al in the hot gas is located in the launch region of outflows through chimneys
 that reach tens of kpc into the halo. Because \al is a good tracer of massive star ejecta in general,
 this has implications for all massive-star ejecta.}
 We first review ejection, propagation and mixing of stellar 
 ejecta in Sect.~\ref{sec:ejectaFate}, then describe the simulation in Sect. ~\ref{sec:sims}
 with results in Sect.~\ref{sec:res}. 
We discuss our results on the background of complementary work
 in Sect. \ref{sec:disc} and summarise our conclusions in Sect.~\ref{sec:conc}.}
 
\section{The path of massive star ejecta}\label{sec:ejectaFate}
Models predict that \al may be ejected
during the main-sequence or Wolf-Rayet phase of a single or binary
massive star \citep{GM12,Brinkmea19}, 
and in supernovae \rvb{\citep{LC06,WH07,NKT13}}.
\rchange{The explosive contribution can be addressed with the isotopic ratio \feu/\alu, where
\fe has a decay time of 4~Myr and is thought to be produced in supernovae, only. \fe is found in deep sea 
sediments, suggesting a relatively nearby supernova 2-3~Myr ago \citep{Feigea18}. 
The non-detection of \al would constrain the 
\feu/\al ratio for this supernova to $>0.18$, well consistent with expectations of $\approx 2$ \citep{Austea17}. 
INTEGRAL observations of the interstellar medium find \feu/\alu=0.2-0.4 \citep{WangWea20}, 
which suggests
a significant or even dominant non-supernova contribution.
}

\begin{figure*}\centering
	\includegraphics[width=0.49\textwidth]{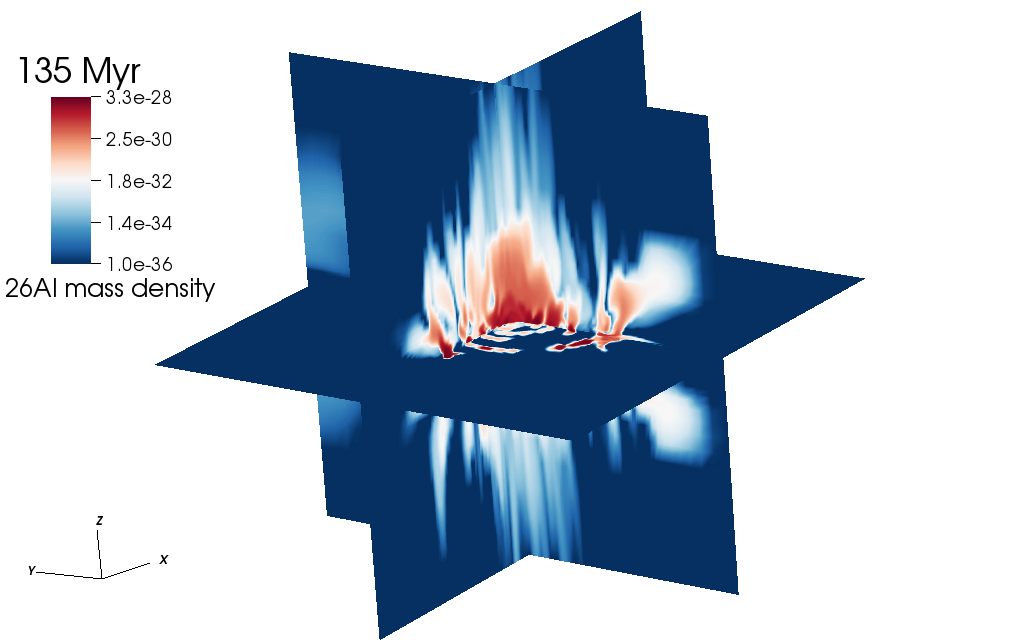}
	\includegraphics[width=0.49\textwidth]{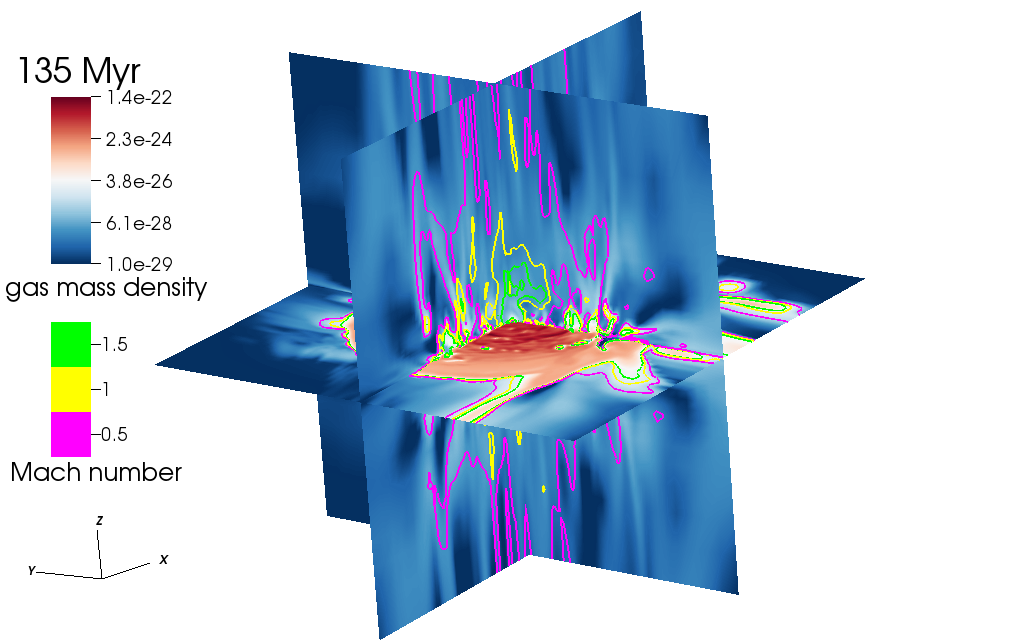}
	\caption{3D renderings of run \rvb{LoH+} at 135~Myr. Both visualisations extend 20~kpc from the centre
	of the simulated galaxy along each axis in positive and negative directions. Left: \al density.
	\al streams off from the disc to significant scale height. Right: Density with contours of the Mach number
	at 0.5, 1 and 1.5. \al marks the base of a larger outflow.
	Movies are available with the online version of the journal. }
    	\label{fig:LoH+-3D}
\end{figure*}

 \begin{figure}\centering
	\includegraphics[width=0.2\textwidth]{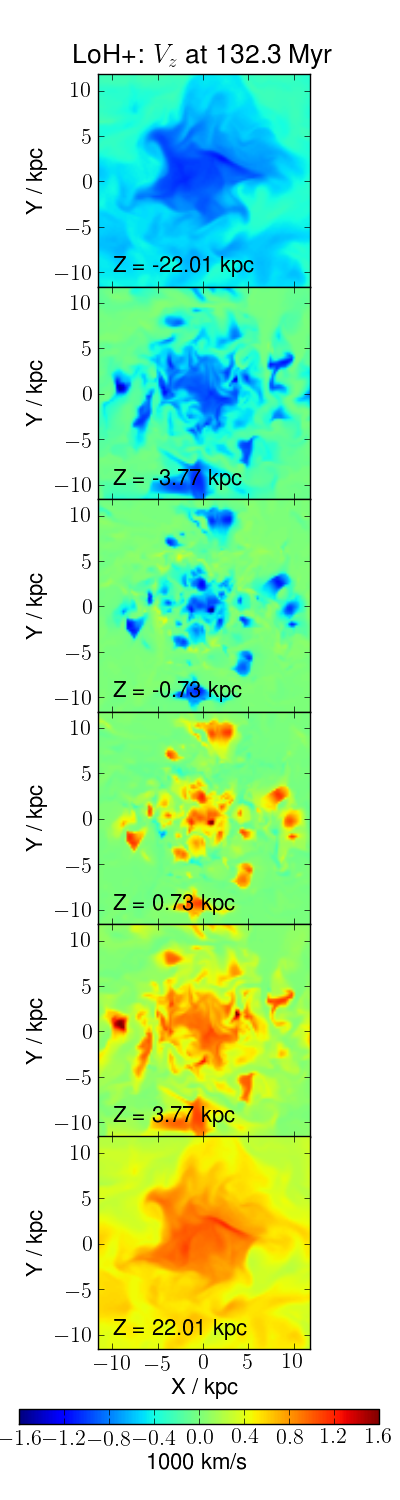}
	\includegraphics[width=0.2\textwidth]{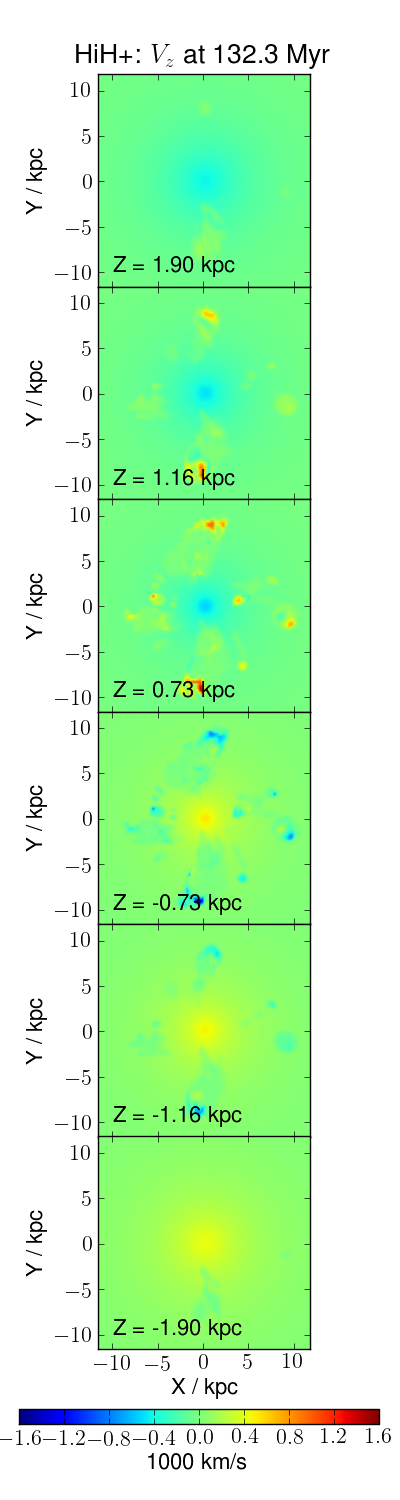}
	\caption{\revI{Halo flow pattern for the two simulations analysed. 
	Shown is the velocity vertical to the disc at different altitudes indicated in the individual panels. 
	Run \rvb{LoH+} (left) with an observationally
	constrained halo density develops fast winds locally, that ascend in chimneys and have turned subsonic 
	by 20~kpc above the disc. This turbulent flow extends beyond the simulation box 
	(100~kpc each way). Run \rvb{HiH+} (right) with higher halo density develops a convection zone
	of $\approx 1-2$~kpc extent around the galactic disc.}
    	\label{fig:vz}}
\end{figure}

Massive stars \rvb{form} in groups and clusters \citep{ZY07}, and quickly ionise their surroundings, producing
parsec-scale HII regions on a timescale of $10^5$~yr
\citep{MackeyJea15}. Massive star winds
then energise and heat the interior of bubbles of the order of 10~pc to X-ray temperatures
$\approx10^6$~yr after star formation
\citep[e.g.][]{Weavea77,Churchea06,EC10,Krausea13a,Krausea14a}.
\rchange{The clustering of massive stars implies the formation of superbubbles
\citep[e.g.][]{OG04,BdA06,Jaskea11,Sasea11,Krausea13a,Krausea18b,Schulrea18a}.}

\rchange{After being expelled from their sources, the massive-star ejecta are shocked to high temperatures within these bubbles
\citep[e.g.,][their Fig.~6]{Krausea13a}. Then they gradually mix trans-sonically, through
turbulence and instabilities, on the sound crossing timescale into the hot gas
throughout the entire superbubble
\citep{Breitschwea16,Krausea18b}. Observations
point to efficient turbulent mixing between hot and cold gas: 
Without such mixing, the superbubbles would be too hot
\citep{DPC01} and would grow too quickly \citep[e.g.][]{OG04}.} High-resolution
3D~hydrodynamics simulations have shown that a turbulent mixing zone between the hot interior
and the shell arises naturally via instabilities due to the time-dependent nature of the energy supply.
This dissipates the bulk of the energy, and ensures realistic bubble growth 
\citep{KD14}. Entrainment of cold shell gas cools the superbubble interior 
to temperatures around or just below 1~keV and
enhances the luminosities, both in good agreement with observations \citep{Krausea14a}. Mixing is strongest just after supernova
explosions, thus explaining the Myr timescale X-ray variability of superbubbles \citep{KD14}.
Other processes that mix massive star ejecta with dense gas include 
ablation of clouds enclosed in the hot bubbles 
\citep{RP13}. Hot, ejecta-enriched gas surrounds such clouds \citep{Gaczkea15,Gaczkea17}, 
squashing them because of pressure differences which may induce the formation of stars \citep{Krausea18b}.

\citet{FKT18} followed the diffusion of \al ejected by massive stars in a 3D~hydrodynamic
simulation
of a whole galaxy and found that the range of inferred isotopic ratios of \al to stable $^{27}$Al
in self-consistently formed stars was in good agreement with values from solar system meteorites,
implying that the Sun was a typical star. They also show that most of their \al is found in the 
cold phase, with 56~per cent of the total mass being in molecular clouds.  \citet{Pleintea19aph}
have compared their simulation results to the gamma-ray data,
and found that the simulation could not account for the high observed scale heights of \al in the Milky Way.
They concluded that star formation in the simulation was likely too fine-grained, possibly due to the neglect
of prominent spiral arms. It appears possible that this may have led to enhanced interaction between the gas phases 
in that simulation, and an overestimate of \al in the star-forming gas.
\rvb{More recently, \citet{FKI20} have followed up on this with new simulations. 
They show that material spiral arms that form spontaneously from self-gravity in the disc also cannot solve 
the scale-height problem and suggest instead that the Sun is at a special place where foreground 
dominates the \al signal. While there is some evidence for material arms, 
the exact nature of the spiral arms of the Milky Way is still debated
\citep[e.g.,][see \citet{FKI20} for more discussion]{Sellwea19,PRS20}.}

\citet{RodgLea19} performed 
similar simulations (details below), but with a 
more coarsly-grained star formation prescription than that of \citet{FKT18}.
\rvb{They imposed spiral arms by an external potential.}
Comparing to
gamma ray observations of \alu, they generally found good agreement, but some discrepancy
regarding the kinematic structure. While the latter cannot be reproduced in detail,
\rchange{possibly because details of star formation in spiral arms are not taken into account accurately enough,}
they show that velocities comparable to the observed ones (different from the cold gas) are present in the simulation.

\section{Simulations}\label{sec:sims}
The simulation we analyse here have been presented in detail in 
\citet{RodgLea19}. In short, it is a set of 3D hydrodynamic simulation\revI{s} of the interstellar medium of 
an isolated disc galaxy set in a gravitational potential that
represents the static dark matter halo, stellar bulge, disc and rotating spiral arms.
Superbubbles representing star forming regions of $10^6 M_\odot$ were injected
at predetermined positions near spiral arms
with energy, mass and \al mass equivalent to a state at 10~Myr after star formation.
The parameters for the gravitational potential, the disc size and the star formation rate 
were chosen to resemble the Milky 
Way\footnote{\rvb{We have become aware of a factor of two error in the energy injection rate, which means that the energy injection in these simulations corresponded to a star formation rate of 6~\mspy rather than 3~\mspy. For the present paper, we have re-run a representative simulation (lower density halo, superbubbles offset towards the leading edge, 'LoH+') with the correct energy input rate. This reduces the scale height of \al somewhat, but otherwise does not lead to a significant change of the results, in particular the predictions for the observed \al kinematics. We use the new LoH+ run in the present paper but keep the original comparison run with high halo density ('HiH+'). }}. 
Standard optically thin cooling down to $10^4$~K
was implemented, where the gas was assumed to be kept ionised by ultraviolet radiation
of stars. Radiative cooling below $10^4$~K was not allowed, but such gas can
still cool via adiabatic expansion. In \citet{RodgLea19} we presented several simulations 
where 
the position of the superbubbles relative to the ridge line of the spiral arms had been varied
as well as the density of the hydrostatic gas halo. For the present analysis, we verified
that the results do not depend on the details of the positioning of the superbubbles with respect
to the spiral arms and we only show results \rvb{from the simulations with the superbubbles offset towards the leading edges of the spiral arms.} 

As we show below, the simulation 
with the observationally constrained lower
central halo gas density of $4.4 \times 10^{-28}$~g~cm$^{-3}$ (run \rvb{LoH+} in the following)
develops a patchy, chimney-like wind structure that transitions to subsonic flow at high altitudes 
above and below the disc (Fig.~\ref{fig:LoH+-3D}), whereas the one with a central density of 
$1.8 \times 10^{-25}$~g~cm$^{-3}$ (run \rvb{HiH+} in the following) develops no large-scale
convection beyond a few kpc.
The halo density in run \rvb{HiH+} is unrealistically high. We include it here to show that
even for this strong variation of the halo density and resulting flow structure, the 
changes to the branching ratio of \al diffusing into, respectively, hot and cold gas are moderate. 

\begin{figure}\centering
	\includegraphics[width=0.32\textwidth]{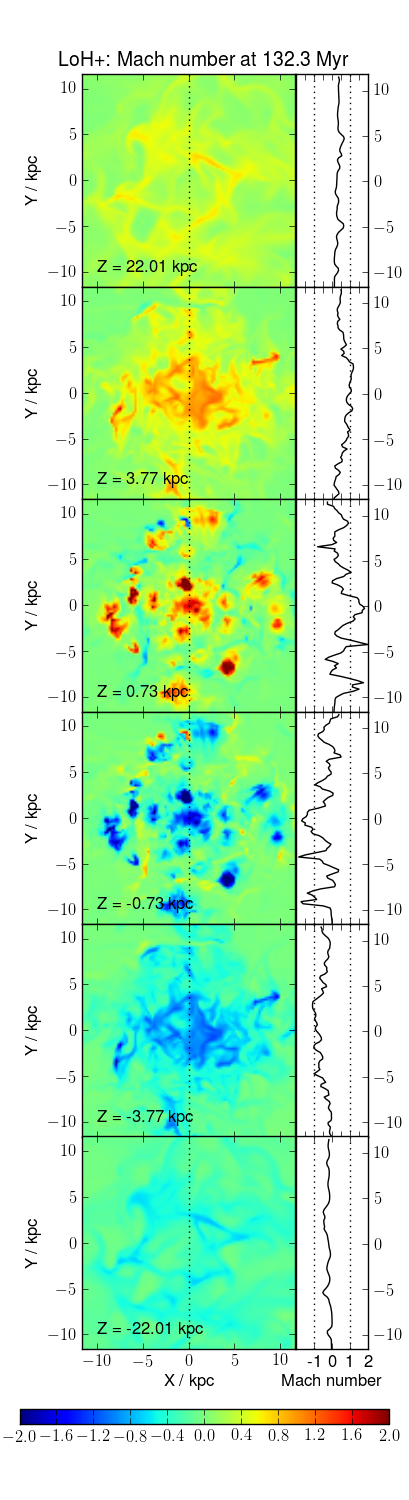}
	\caption{\revI{Maps and cross sections of the vertical Mach number for various slices above and below
	the disc midplane for run \rvb{LoH+}. The altitude is given in the individual panels. 
	A dotted line in the maps at each altitude denotes the location for the cross section shown 
	on the right of each map. In the latter plots, two dashed lines indicate Mach numbers of 
	one in upward (positive) and downward (negative) directions, respectively.}
    	\label{fig:Mz}}
\end{figure}

\begin{figure}\centering
	\includegraphics[width=0.47\textwidth]{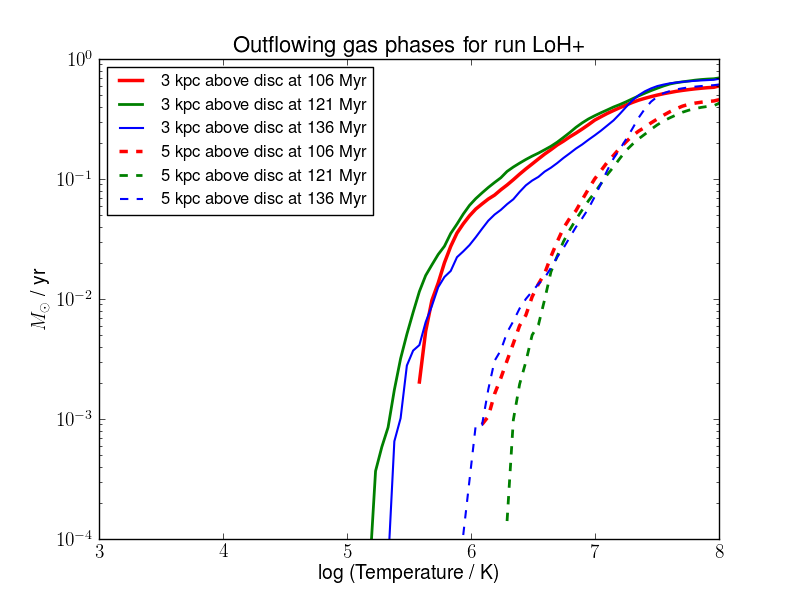}
		\caption{\rvb{Cumulative temperature histograms weighted with the outflowing mass flux for a
	circular cross section of  9~kpc radius 3 and 5 kpc above the disc for run LoH+. 
		Only outflowing gas is taken into account.
	We find no outflowing gas below $10^5$~K. Gas with $10^5-10^6$~K falls back
	to the disc before reaching 5~kpc altitude. From then on, only gas with at least $10^6$K is present
	in the outflow. }
    		\label{fig:outflowphases}
    	}	

\end{figure}

\begin{figure*}\centering
	\includegraphics[width=0.47\textwidth]{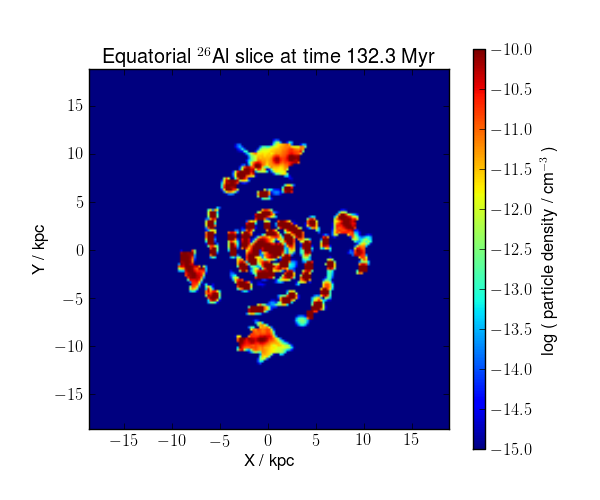}
	\includegraphics[width=0.47\textwidth]{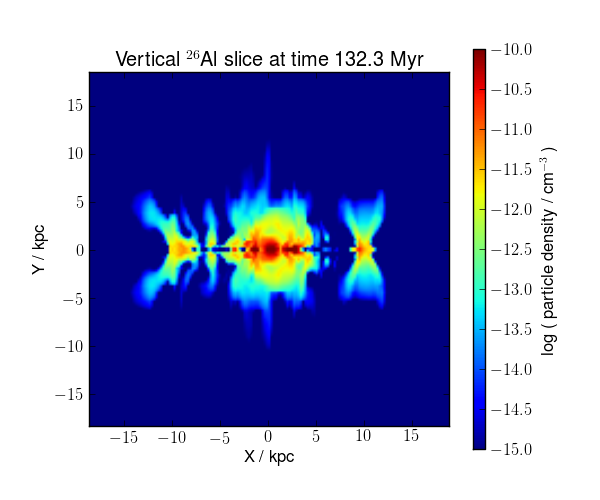}
	\caption{Slice through the equatorial midplane (left) and a vertical midplane (right)
	\revI{for run \rvb{LoH+}} at 132.3~Myr. 
	We show the density of \al nuclei. \al  fills long-lasting cavities with repeated 
	cycles of injection, advection out of the galactic plane and decay. A movie is provided with the
	online version of the journal.
    	\label{fig:simslice}}
\end{figure*}


\begin{figure*}\centering
	\includegraphics[width=0.47\textwidth]{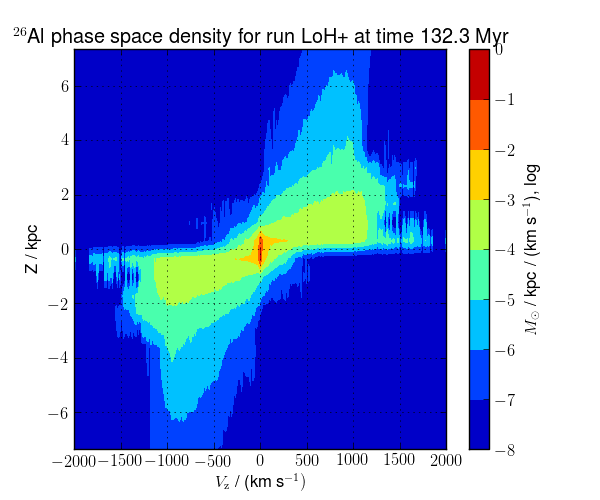}
	\includegraphics[width=0.47\textwidth]{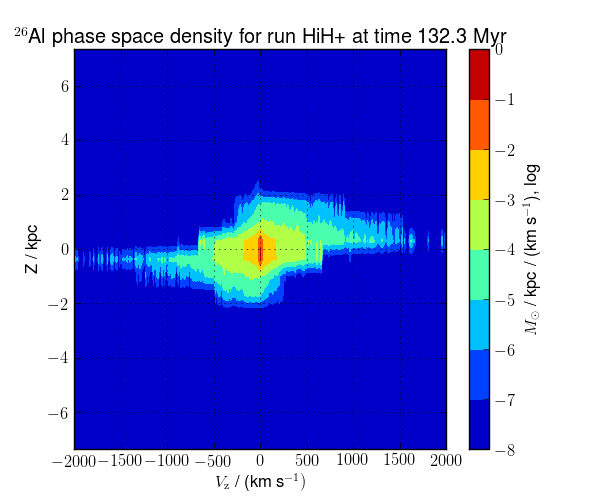}
	\caption{\revI{Phase space density for \al  for runs \rvb{LoH+} (left) and \rvb{HiH+} (right).
	Shown are 2D distributions of the coordinate vertical to the disc and the velocity in the same direction.
	 \al is confined by the high density halo gas in run \rvb{HiH+}.}
    	\label{fig:phasespace}}
\end{figure*}

\section{Results}\label{sec:res}

\revI{We demonstrate the difference for the flow pattern of the two simulations in Fig.~\ref{fig:vz}.
There are two basic hydrodynamic flow patterns for galaxy halos, supersonic winds
\citep{CC85} and lower halo convection \citep{dAB04}. The observational threshold for formation
of winds in star-forming galaxies is around a star formation rate density of 
$0.1 M_\odot$~yr$^{-1}$~kpc$^{-1}$
\citep{Heckmea15}. This agrees well
with the onset of winds in similar simulations where stellar feedback is implemented in
the form of superbubbles \citep{vGlea13}. \rvb{We have used a 
star formation rate of $3\, (6) M_\odot$~yr$^{-1}$ for run LoH+ (HiH+)}. So, averaged over the whole disc
the star formation rate density is $0.01 \,\rvb{(0.02)} M_\odot$~yr$^{-1}$~kpc$^{-1}$, well below
the threshold for wind formation. Locally, for each implemented superbubble, 
the star formation rate density is $0.04 \,\rvb{(0.08)} M_\odot$~yr$^{-1}$~kpc$^{-1}$.
This comes close enough to the threshold, so that marginally supersonic
outflows are launched in our simulations locally from each superbubble (Fig.~\ref{fig:Mz}).
\rvb{These outflows contain hot gas $\gtrsim 10^6$~K, only (Fig.~\ref{fig:outflowphases}), and 
would therefore not be expected to show up in ultraviolet wind studies such as the one of
\citep{Heckmea15}.}
In run \rvb{LoH+}, these outflows rise conically, vertically away from the disc.
About 20~kpc above and below the disc, the cones have merged and decelerated to
a subsonic flow. The flow then streams out of our simulation box
at 100~kpc. In run \rvb{HiH+}, the high density of the halo chokes the outflows and the circulation
is restricted to a $\approx 2$~kpc region around the disc (Fig.~\ref{fig:vz}, right).
}

\revI{We show midplane slices of the density of \al nuclei in Fig~\ref{fig:simslice} for run \rvb{LoH+}.
A movie is provided with the online version of the journal. The superbubbles form long-lasting
cavities that are repeatedly filled with \alu-rich ejecta. They are advected into the halo in chimneys,
where radioactive decay limits the scale height achieved by the ejecta. This can be seen
more clearly in the two-dimensional slices of the phase space density of the \al nuclei shown in
Fig.~\ref{fig:phasespace}. We show the $z-v_z$-slice, where $z$ denotes the vertical coordinate,
for both simulations. For run \rvb{LoH+}, we find large outflow velocities in excess of 1000~km/s for the \al nuclei.
At these velocities, \al advances by 1-2~kpc per decay time. The decay is clearly visible along
the inner 7~kpc of the outflow. In contrast, the ejecta in run \rvb{HiH+} are limited by the dense halo.
Only a small fraction of the ejecta acquires velocities above 500~km/s, and only near the launch point
of the outflows.  }

\begin{figure*}\centering
	\includegraphics[width=0.47\textwidth]{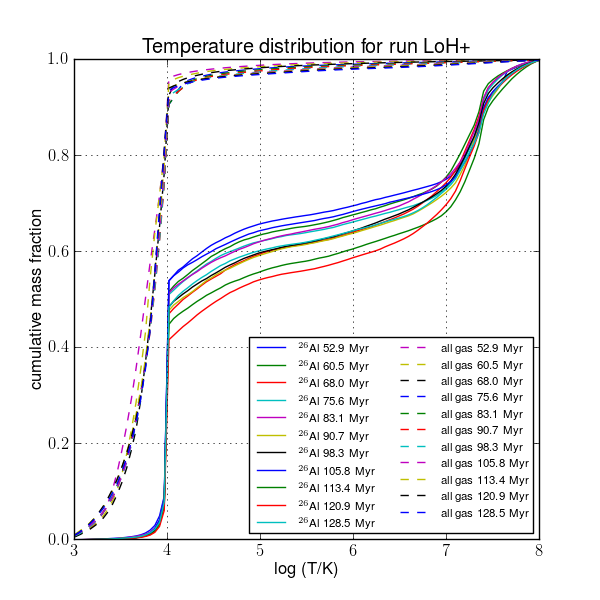}
	\includegraphics[width=0.47\textwidth]{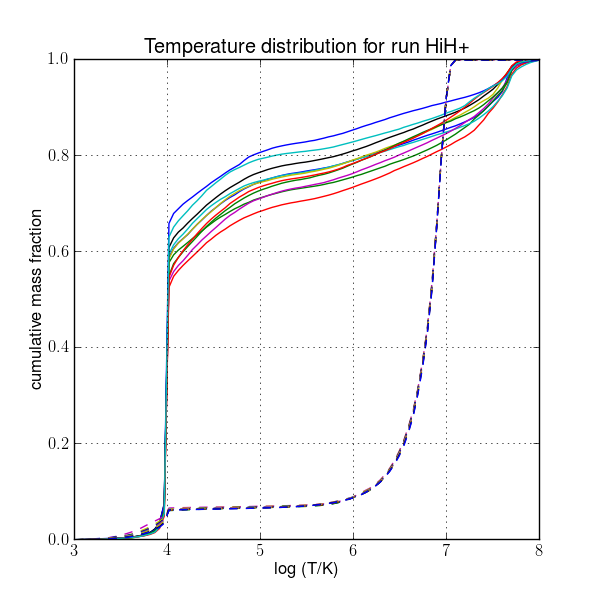}
	\caption{Temperature distributions of \al (solid lines) and all gas 
	(dashed lines) for typical snapshots \revI{for run \rvb{LoH+} (left) and run \rvb{HiH+} (right), respectively}.
	The distributions \revI{for \al} differ significantly \revI{from the ones for all gas in both cases.	
	}	
	 with about half of the \al being found 
	in hot gas. The lines do not form a \revI{monotonic} sequence in time, i.e., the plot shows 
	representative random fluctuations of the distribution.
    	\label{fig:Tdist}}
\end{figure*}

The temperature distribution by mass fraction for \al is compared to the 
one for all gas for representative snapshot times in Fig.~\ref{fig:Tdist}.
The distribution for \al has some variation in time. Overall, there is
a marked difference to the general gas distribution. 
\revI{This is true for the low density run \rvb{LoH+}, where the hot halo (inside our simulated box)
contributes only a few per cent of the overall baryonic mass, as well as in run \rvb{HiH+},
where it contributes around 90 per cent. In run \rvb{LoH+},
}
\rvb{a significant fraction} of the \al mass remains at high temperatures with typically
\rvb{30-}40~per cent above $10^6$~K.
\revI{This fraction decreases to 20 per cent for run \rvb{HiH+}.}

\begin{figure*}\centering
	\includegraphics[width=0.47\textwidth]{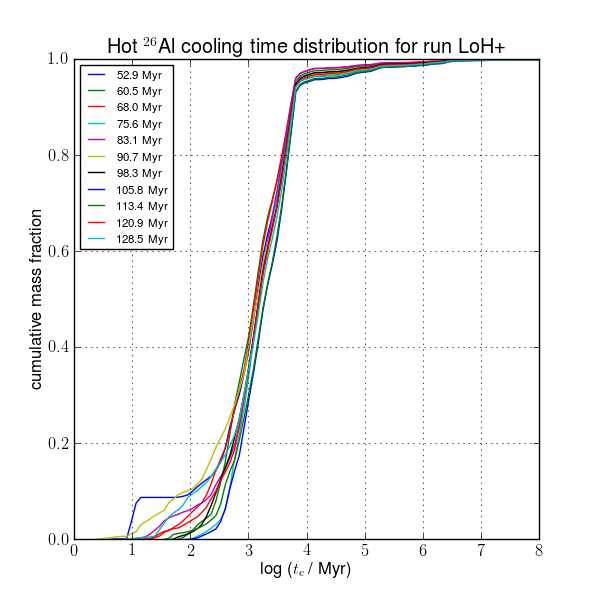}
	\includegraphics[width=0.47\textwidth]{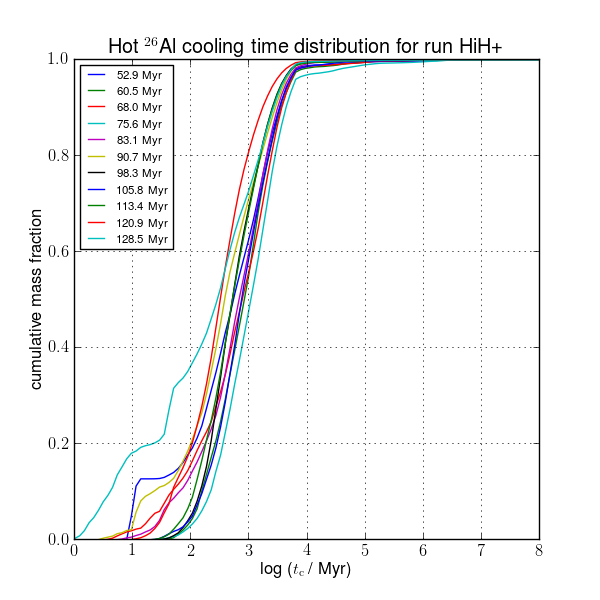}
	\caption{Cooling time distribution for \al carrying hot gas ($T>10^6$K) 
	assuming solar metallicity for representative snapshots.
	Shown is the cumulative \al mass over the cooling time.
	Almost all hot gas cooling times exceed the radioactive decay time of \al ($\approx 1$~Myr)
	significantly. 
    	\label{fig:tcdist}}
\end{figure*}

The \al in the cold component is likely to quickly find its way into new stars.
For the hot one, a typical timescale is given by the radiative cooling timescale.
To obtain a lower limit for the radiative cooling time $t_\mathrm{c}$, 
we assume solar metallicity and use
the collisional ionisation equilibrium cooling curve of \citet{SD93}, $\Lambda(T)$,
$t_\mathrm{c} = {k_\mathrm{B}T}/{(n\Lambda)}$,
where $k_\mathrm{B}$ is the Boltzmann constant and $n$ is the particle density.
We show the cooling time distributions for hot gas with temperature $T>10^6$K
that carries \al in Fig~\ref{fig:tcdist}.
The median for run \rvb{LoH+} is a few Gyr. Less then about
10~per cent of the \al mass is associated with gas that can cool on the dynamical timescale
of a galaxy ($\sim 100$~Myr). 
\revI{For run \rvb{HiH+}, the cooling time distribution becomes more variable. The median is typically
just below a Gyr, and the fraction of \al-traced gas that can cool on a dynamical timescale
of the galaxy varies between a few and above 30~per cent.}

\section{Discussion}\label{sec:disc}


\revI{Our simulations do not include the first 10~Myr of star formation.
This is a reasonable approximation for the study of larg\rvb{e}-scale interstellar medium dynamics,
because it is well-known that the embedded phase of star formation lasts at most a few Myr,
and that most of the \al (and other metals) are ejected into large superbubbles. This
is especially true for supernova ejecta.}

\revI{It is interesting in this context to compare our results to \citet{FKT18}, who
model the physics of star formation in much more detail,
but do not include spiral arms in their simulations. They find that}
a significant fraction of \al ends up in the next generation of
stars, being mixed on the \al decay timescale. 
\rchange{In typical stars, the predicted isotopic ratio 
\alu/$^{27}$Al is therefore reduced by about one order of magnitude, only, from the average 
interstellar medium value, which had been determined from gamma-ray surveys to be  $\approx 5\times 10^{-4}$\citep{Diehl13}.
The Sun, with an initial value of $\approx 5\times 10^{-5}$ would 
therefore be a rather typical star. }
Our simulation makes very different assumptions about the spatial arrangement
of stellar feedback. We impose spiral arms, ignore the first 10~Myr after star formation, 
and instead inject superbubbles at this stage with a diameter of 600 pc. 
\rchange{We} still find in our simulation results that \revI{at least} half of the \al by mass diffuses into the cold phase.
\rchange{The agreement between the two simulations suggests that \revI{for the scales studied here,} mixing does not strongly depend 
on small-scale details unresolved in our simulation\revI{s} and that we capture mixing reasonably well.}

\revI{For our main run \rvb{LoH+},} \al in the hot phase  traces a fraction of about \rvb{30-}40~per cent of the 
massive star ejecta that diffuse out of the galaxy into its gaseous halo.
The outflow \rvb{is hot (Fig.~\ref{fig:outflowphases}) and} can be seen directly in \revI{Figs.~\ref{fig:LoH+-3D}, \ref{fig:simslice} and \ref{fig:phasespace}}. Typical cooling
times are of the order of Gyr. This identifies
a component of interstellar gas that characterises the enrichment of the circumgalactic and possibly intergalactic
medium.
\rvb{Even in} the simulation with unrealistically high halo density (\rvb{HiH+}), 
this fraction is reduced by a factor of two\rvb{, only}.
It thus gives us a robust limit for the branching ratio of \al in hot and cold gas at the time of its decay.

\rvb{The occurence of the outflow depends strongly on the assumed halo density. In contrast to the higher halo density simulation, the lower halo density run produced the outflow even though it had a lower star-formation rate. The outflow occurs somewhat below the observational threshold for the star-formation rate surface density from winds detected in the ultraviolet band (compare Sect.~\ref{sec:res}). Our outflow is, however, too hot to be detected in this way and would instead require sensitive high-resolution spectroscopy in X-rays to be observed. }

\begin{figure}\centering
	\includegraphics[width=0.49\textwidth]{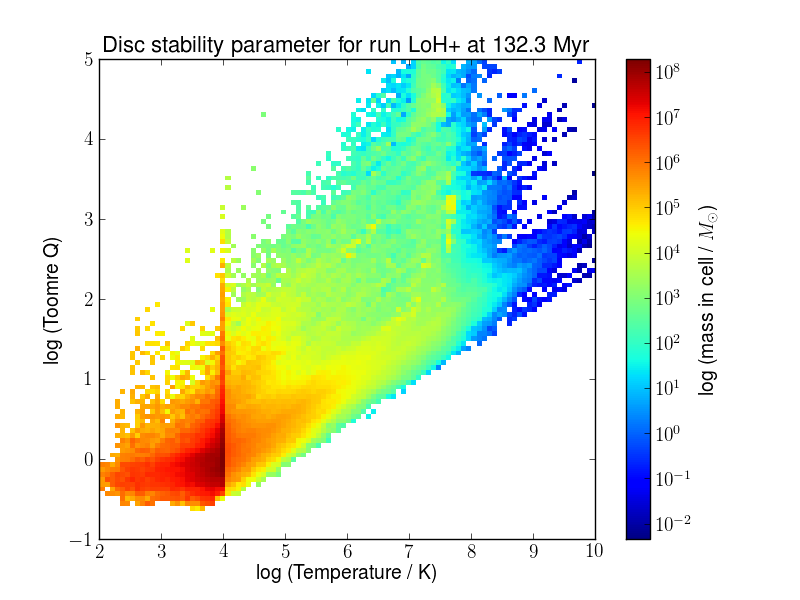}
	\caption{\rvb{Two-dimensional, mass-weighted histogram of temperature
	and Toomre Q parameter towards the end of run LoH+ for the disc region. 
	The cool gas generally accumulates around or
	just below $Q=1$, as expected.}
    	\label{fig:dstab}}
\end{figure}

\begin{figure}\centering
	\includegraphics[width=0.47\textwidth]{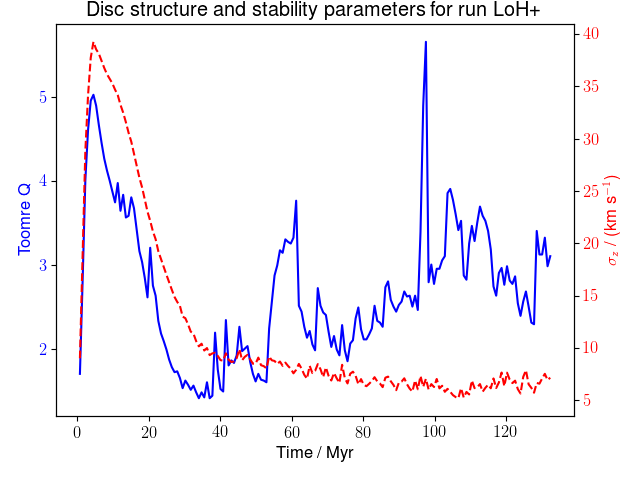}
	\caption{\rvb{Run of mass weighted Toomre~Q parameter and vertical velocity
	dispersion, $\sigma_z$ over time, both for the disc region of run LoH+.} 
    	\label{fig:dstab-time}}
\end{figure}

\rvb{Apart from this hot outflow from our inter-arm regions, the interstellar medium dynamics in our simulations are similar to other theoretical work and observations of 
disc galaxies. We show this here with the 
velocity dispersion and the \citet{Toomre64} $Q$ parameter, which have been used in the literature
to characterise
structure and dynamics of the interstellar medium in disc galaxies \citep[e.g.,][]{Krumhea18,Orrea20}. }
\rvb{\citet{Krumhea18} show that the Toomre stability parameter may
be expressed as:
\begin{equation*}
Q = \frac{f \kappa \sigma}{\pi G \Sigma_\mathrm{g}}\, ,
\end{equation*} 
where $\sigma=\sqrt{3\sigma_z^2+c_\mathrm{s}^2}$ represents the total gas velocity dispersion 
including bulk and thermal motions. $c_\mathrm{s}$ is the speed of sound,
$\kappa$ the epicyclic frequency, $\Sigma_\mathrm{g}$ the gas surface density and 
$f$ a factor that takes into account that the gas surface density rather than the 
total gravitating density is used. We follow \citet{Krumhea18} and adopt $f=0.5$.
$Q\approx 1$ is expected for  a star-forming disc galaxy. 
Figure~\ref{fig:dstab} displays the distribution of the Toomre~Q parameter for the disc region
(cylindrical radius less than 9~kpc, altitude above (below) the disc less than 0.5~kpc)
for run LoH+ towards the end of the simulation as a function of gas temperature. Most of the cold material
indeed assembles around $Q\approx 1$. We show the run of the mass-weighted $Q$-parameter over time in 
Fig.~\ref{fig:dstab-time}. $Q$ is mostly between two and three for the part of the simulation we evaluate.}

\rvb{We also calculate 
the mass-weighted line-of-sight velocity dispersion vertical to the disc, $\sigma_z$, 
for the same disc region as before. It is also plotted against time in Fig.~\ref{fig:dstab-time}. 
For the late times that we are interested in here, the velocity dispersion has converged to about
7~\kms.
This can be compared
to dense gas tracers for which recent simulations and observations find values in the range of 5-30~\kms
\citep{DasMea20,Orrea20}. 
} 

\rvb{We conclude that the parameters that characterise the dynamics of the 
interstellar medium in our simulated galaxies are, overall, realistic. In particular, there is no evidence for
enhanced vertical motion of dense gas that would conflict with observations of star-forming galaxies.}

\cchange{
Over cosmological timescales, if the metals accumulate in the halo, the cooling time would continuously decrease until
the gas would eventually cool and form stars. However, halo gas can also be lost from a galaxy, for example
because of Active Galactic Nucleus (AGN) feedback \citep{TK15}. In particular, similar to processes 
believed to take place in galaxy clusters, radio AGN can produce buoyant bubbles that can drag a large 
fraction of inner halo gas in their wake \citep*{HKA07}. Despite radio AGN being principally 
observed in galaxy clusters they could be equally abundant in all galaxies, which would be in good agreement
with recent LOFAR source counts \citep{Krausea19b}. Much of the halo gas could also be stripped
when the Milky Way will merge into the Virgo cluster \citep{Tulea14}, perhaps reinforced by simultaneous
AGN triggering \citep{Marshea18}.
}

\cchange{Finding the metals to be distributed in roughly equal parts in cold gas and, respectively, in hot
slowly cooling gas approximately 1~Myr after their ejection is in line with the extragalactic data.
Observations of galaxies and galaxy clusters suggest that up to about 50~per cent of the metals
remain in the galaxies \citep{RA14,MM19}. Chemodynamical simulations find a higher 
\revI{retained metal} fraction in
galaxies of higher mass, reaching 80-90~per cent for galaxies with virial masses around $10^{12} M_\odot$,
like the Milky Way. Over cosmological timescales, some of the halo gas in the Milky Way will indeed likely cool and be re-accreted to the disc to again take part in star formation, and only a part of the enriched 
halo gas will escape altogether from the dark-matter halo of the Milky Way.}

\section{Summary and conclusions}\label{sec:conc}
We have analysed a simulation of a Milky-Way-like galaxy
that traces radioactive \al through different gas phases. 
\al is a unique tracer for massive star ejecta, 
as it provides an additional gauge
against, on one hand, INTEGRAL observations of the 1.8~MeV 
radioactive decay line throughout the Galaxy, and, on the other hand, 
against solar-system data from deposits of recent traces of massive-star ejecta and
meteoritic isotopic abundance measurements for the early \rvb{Solar System}.

We find from our simulations that \revI{for a Milky-Way-like galaxy
with superbubbles concentrated towards the spiral arms}
about \rvb{30-}40~per cent of the \al traced massive-star ejecta
are ejected out into the gaseous halo at high temperatures.
\revI{Even when we choke off the outflow with an unrealistically high halo density,
we still get 20~per cent of the \al into gas with temperatures above $10^6$~K.}
Cooling times for this component are of the order of Gyr.
These ejecta are therefore not immediately available for increasing the metallicity in the stellar
population. The rest has diffused into gas of low temperature near our cooling limit.

These findings are in good agreement with gamma-ray data requiring
a high fraction of the Galactic \revI{radioactive} \al output to remain in hot gas.
\revI{We also find a high fraction of \al in cold gas. We do not include the formation of stars in our simulations.
But in view of the results of  \citet{FKT18}, who also performed galaxy-scale simulations
of \rvb{the} interstellar medium, but focussed on the diffusion of \al into newly formed stars,
a consistent view emerges: for galaxies similar to the Milky Way,
metals produced by massive stars seem to branch in 
\rvb{similar} parts into cold, star forming gas and hot gas venting into the galaxy halo.}

Chemodynamical cosmological simulations would require a higher fraction of metals 
to get into stars for galaxies the size of the Milky Way over cosmological timescales.
\revI{This may indicate that over much of its lifetime, these galaxies
have star formation more evenly spread out in the disc and less concentrated towards the spiral arms.
This would confine the outflows and likely enhance metal mixing into the cold gas, as demonstrated
in run \rvb{HiH+} where we confined the outflow with a high halo density. Alternatively, t}his
may point to  convective diffusion of these metals back into the Galaxy on such timescales.
Overall, \rvb{we find a comparable amount (within a factor of a few)} of freshly produced metals in hot and cold gas 
\rvb{respectively} in our simulations,
\rvb{which}
appears remarkably similar to the ratio of metals in, respectively, stars and gas, 
inferred from galaxy clusters and chemical evolution modelling of galaxies.

\section*{Acknowledgements}
This work has made use of the University of Hertfordshire's high-performance computing facility.
\rvb{C.K. acknowledges funding from the UK Science and Technology Facility Council (STFC) through grant ST/M000958/1 \& ST/ R000905/1. DRL acknowledges funding from the European Research Council (ERC) under the European Union's Horizon 2020 research and innovation programme (grant agreement No 817540, ASTROFLOW). 
This article is based upon work from the “ChETEC” COST Action (CA16117), supported by COST (European Cooperation in Science and Technology).
Figure~\ref{fig:LoH+-3D} was produced using the visualisation tool VisIt \citep{Childsea12}. We thank the anonymous referee for useful comments that helped to improve the manuscript.
}

\section*{Data availability}
Data available on request.


\bibliographystyle{mnras}
\bibliography{/Users/mghkrause/texinput/references}

\begin{thebibliography}{}
\makeatletter
\relax
\def\mn@urlcharsother{\let\do\@makeother \do\$\do\&\do\#\do\^\do\_\do\%\do\~}
\def\mn@doi{\begingroup\mn@urlcharsother \@ifnextchar [ {\mn@doi@}
  {\mn@doi@[]}}
\def\mn@doi@[#1]#2{\def\@tempa{#1}\ifx\@tempa\@empty \href
  {http://dx.doi.org/#2} {doi:#2}\else \href {http://dx.doi.org/#2} {#1}\fi
  \endgroup}
\def\mn@eprint#1#2{\mn@eprint@#1:#2::\@nil}
\def\mn@eprint@arXiv#1{\href {http://arxiv.org/abs/#1} {{\tt arXiv:#1}}}
\def\mn@eprint@dblp#1{\href {http://dblp.uni-trier.de/rec/bibtex/#1.xml}
  {dblp:#1}}
\def\mn@eprint@#1:#2:#3:#4\@nil{\def\@tempa {#1}\def\@tempb {#2}\def\@tempc
  {#3}\ifx \@tempc \@empty \let \@tempc \@tempb \let \@tempb \@tempa \fi \ifx
  \@tempb \@empty \def\@tempb {arXiv}\fi \@ifundefined
  {mn@eprint@\@tempb}{\@tempb:\@tempc}{\expandafter \expandafter \csname
  mn@eprint@\@tempb\endcsname \expandafter{\@tempc}}}

\bibitem[\protect\citeauthoryear{{Austin}, {West}  \& {Heger}}{{Austin}
  et~al.}{2017}]{Austea17}
{Austin} S.~M.,  {West} C.,   {Heger} A.,  2017, \mn@doi [\apjl]
  {10.3847/2041-8213/aa68e7}, \href
  {https://ui.adsabs.harvard.edu/abs/2017ApJ...839L...9A} {839, L9}

\bibitem[\protect\citeauthoryear{{B{\"o}hringer} \& {Werner}}{{B{\"o}hringer}
  \& {Werner}}{2010}]{BW10}
{B{\"o}hringer} H.,  {Werner} N.,  2010, \mn@doi [\aapr]
  {10.1007/s00159-009-0023-3}, \href
  {http://cdsads.u-strasbg.fr/abs/2010A%26ARv..18..127B} {18, 127}

\bibitem[\protect\citeauthoryear{{Bouchet}, {Jourdain}  \& {Roques}}{{Bouchet}
  et~al.}{2015}]{BJR15}
{Bouchet} L.,  {Jourdain} E.,   {Roques} J.-P.,  2015, \mn@doi [\apj]
  {10.1088/0004-637X/801/2/142}, \href
  {http://adsabs.harvard.edu/abs/2015ApJ...801..142B} {801, 142}

\bibitem[\protect\citeauthoryear{{Breitschwerdt} \& {de
  Avillez}}{{Breitschwerdt} \& {de Avillez}}{2006}]{BdA06}
{Breitschwerdt} D.,  {de Avillez} M.~A.,  2006, \mn@doi [\aap]
  {10.1051/0004-6361:20064989}, \href
  {http://adsabs.harvard.edu/abs/2006A%26A...452L...1B} {452, L1}

\bibitem[\protect\citeauthoryear{{Breitschwerdt}, {Feige}, {Schulreich},
  {Avillez}, {Dettbarn}  \& {Fuchs}}{{Breitschwerdt}
  et~al.}{2016}]{Breitschwea16}
{Breitschwerdt} D.,  {Feige} J.,  {Schulreich} M.~M.,  {Avillez} M.~A.~D.,
  {Dettbarn} C.,   {Fuchs} B.,  2016, \mn@doi [\nat] {10.1038/nature17424},
  \href {http://cdsads.u-strasbg.fr/abs/2016Natur.532...73B} {532, 73}

\bibitem[\protect\citeauthoryear{{Brinkman}, {Doherty}, {Pols}, {Li},
  {C{\^o}t{\'e}}  \& {Lugaro}}{{Brinkman} et~al.}{2019}]{Brinkmea19}
{Brinkman} H.~E.,  {Doherty} C.~L.,  {Pols} O.~R.,  {Li} E.~T.,  {C{\^o}t{\'e}}
  B.,   {Lugaro} M.,  2019, \mn@doi [\apj] {10.3847/1538-4357/ab40ae}, \href
  {https://ui.adsabs.harvard.edu/abs/2019ApJ...884...38B} {884, 38}

\bibitem[\protect\citeauthoryear{{Chevalier} \& {Clegg}}{{Chevalier} \&
  {Clegg}}{1985}]{CC85}
{Chevalier} R.~A.,  {Clegg} A.~W.,  1985, \mn@doi [\nat] {10.1038/317044a0},
  \href {http://adsabs.harvard.edu/abs/1985Natur.317...44C} {317, 44}

\bibitem[\protect\citeauthoryear{Childs et~al.,}{Childs
  et~al.}{2012}]{Childsea12}
Childs H.,  et~al., 2012, {High Performance Visualization--Enabling
  Extreme-Scale Scientific Insight}, pp 357--372

\bibitem[\protect\citeauthoryear{{Churchwell} et~al.,}{{Churchwell}
  et~al.}{2006}]{Churchea06}
{Churchwell} E.,  et~al., 2006, \mn@doi [\apj] {10.1086/507015}, \href
  {http://adsabs.harvard.edu/abs/2006ApJ...649..759C} {649, 759}

\bibitem[\protect\citeauthoryear{{Das}, {McGaugh}, {Ianjamasimanana},
  {Schombert}  \& {Dwarakanath}}{{Das} et~al.}{2020}]{DasMea20}
{Das} M.,  {McGaugh} S.~S.,  {Ianjamasimanana} R.,  {Schombert} J.,
  {Dwarakanath} K.~S.,  2020, \mn@doi [\apj] {10.3847/1538-4357/ab5fcd}, \href
  {https://ui.adsabs.harvard.edu/abs/2020ApJ...889...10D} {889, 10}

\bibitem[\protect\citeauthoryear{{Diehl}}{{Diehl}}{2013}]{Diehl13}
{Diehl} R.,  2013, \mn@doi [Reports on Progress in Physics]
  {10.1088/0034-4885/76/2/026301}, \href
  {http://cdsads.u-strasbg.fr/abs/2013RPPh...76b6301D} {76, 026301}

\bibitem[\protect\citeauthoryear{{Diehl} et~al.,}{{Diehl}
  et~al.}{1995}]{Diehlea95}
{Diehl} R.,  et~al., 1995, \aap, \href
  {http://adsabs.harvard.edu/abs/1995A%26A...298..445D} {298, 445}

\bibitem[\protect\citeauthoryear{{Diehl} et~al.,}{{Diehl}
  et~al.}{2006}]{Diehlea06}
{Diehl} R.,  et~al., 2006, \mn@doi [\nat] {10.1038/nature04364}, \href
  {http://cdsads.u-strasbg.fr/abs/2006Natur.439...45D} {439, 45}

\bibitem[\protect\citeauthoryear{{Diehl} et~al.,}{{Diehl}
  et~al.}{2014}]{Diehlea14a}
{Diehl} R.,  et~al., 2014, \mn@doi [Science] {10.1126/science.1254738}, \href
  {http://adsabs.harvard.edu/abs/2014Sci...345.1162D} {345, 1162}

\bibitem[\protect\citeauthoryear{{Dunne}, {Points}  \& {Chu}}{{Dunne}
  et~al.}{2001}]{DPC01}
{Dunne} B.~C.,  {Points} S.~D.,   {Chu} Y.-H.,  2001, \mn@doi [\apjs]
  {10.1086/321794}, \href {http://cdsads.u-strasbg.fr/abs/2001ApJS..136..119D}
  {136, 119}

\bibitem[\protect\citeauthoryear{{Dwarkadas}, {Dauphas}, {Meyer}, {Boyajian}
  \& {Bojazi}}{{Dwarkadas} et~al.}{2017}]{Dwarkea17}
{Dwarkadas} V.~V.,  {Dauphas} N.,  {Meyer} B.,  {Boyajian} P.,   {Bojazi} M.,
  2017, \mn@doi [\apj] {10.3847/1538-4357/aa992e}, \href
  {https://ui.adsabs.harvard.edu/abs/2017ApJ...851..147D} {851, 147}

\bibitem[\protect\citeauthoryear{{Everett} \& {Churchwell}}{{Everett} \&
  {Churchwell}}{2010}]{EC10}
{Everett} J.~E.,  {Churchwell} E.,  2010, \mn@doi [\apj]
  {10.1088/0004-637X/713/1/592}, \href
  {http://adsabs.harvard.edu/abs/2010ApJ...713..592E} {713, 592}

\bibitem[\protect\citeauthoryear{{Feige} et~al.,}{{Feige}
  et~al.}{2018}]{Feigea18}
{Feige} J.,  et~al., 2018, \mn@doi [\prl] {10.1103/PhysRevLett.121.221103},
  \href {https://ui.adsabs.harvard.edu/abs/2018PhRvL.121v1103F} {121, 221103}

\bibitem[\protect\citeauthoryear{{Fujimoto}, {Krumholz}  \&
  {Tachibana}}{{Fujimoto} et~al.}{2018}]{FKT18}
{Fujimoto} Y.,  {Krumholz} M.~R.,   {Tachibana} S.,  2018, \mn@doi [\mnras]
  {10.1093/mnras/sty2132}, \href
  {https://ui.adsabs.harvard.edu/abs/2018MNRAS.480.4025F} {480, 4025}

\bibitem[\protect\citeauthoryear{{Fujimoto}, {Krumholz}  \&
  {Inutsuka}}{{Fujimoto} et~al.}{2020}]{FKI20}
{Fujimoto} Y.,  {Krumholz} M.~R.,   {Inutsuka} S.-i.,  2020, \mn@doi [\mnras]
  {10.1093/mnras/staa2125}, \href
  {https://ui.adsabs.harvard.edu/abs/2020MNRAS.497.2442F} {497, 2442}

\bibitem[\protect\citeauthoryear{{Gaczkowski} et~al.,}{{Gaczkowski}
  et~al.}{2015}]{Gaczkea15}
{Gaczkowski} B.,  et~al., 2015, \mn@doi [\aap] {10.1051/0004-6361/201526527},
  \href {http://cdsads.u-strasbg.fr/abs/2015A%26A...584A..36G} {584, A36}

\bibitem[\protect\citeauthoryear{{Gaczkowski} et~al.,}{{Gaczkowski}
  et~al.}{2017}]{Gaczkea17}
{Gaczkowski} B.,  et~al., 2017, \mn@doi [\aap] {10.1051/0004-6361/201628508},
  \href {http://adsabs.harvard.edu/abs/2017A%26A...608A.102G} {608, A102}

\bibitem[\protect\citeauthoryear{{Gounelle}}{{Gounelle}}{2015}]{Gounea15}
{Gounelle} M.,  2015, \mn@doi [\aap] {10.1051/0004-6361/201526174}, \href
  {https://ui.adsabs.harvard.edu/abs/2015A&A...582A..26G} {582, A26}

\bibitem[\protect\citeauthoryear{{Gounelle} \& {Meynet}}{{Gounelle} \&
  {Meynet}}{2012}]{GM12}
{Gounelle} M.,  {Meynet} G.,  2012, \mn@doi [\aap]
  {10.1051/0004-6361/201219031}, \href
  {http://cdsads.u-strasbg.fr/abs/2012A%26A...545A...4G} {545, A4}

\bibitem[\protect\citeauthoryear{{Groopman} et~al.,}{{Groopman}
  et~al.}{2015}]{Groopmea15}
{Groopman} E.,  et~al., 2015, \mn@doi [\apj] {10.1088/0004-637X/809/1/31},
  \href {https://ui.adsabs.harvard.edu/abs/2015ApJ...809...31G} {809, 31}

\bibitem[\protect\citeauthoryear{{Heath}, {Krause}  \& {Alexander}}{{Heath}
  et~al.}{2007}]{HKA07}
{Heath} D.,  {Krause} M.,   {Alexander} P.,  2007, \mn@doi [\mnras]
  {10.1111/j.1365-2966.2006.11191.x}, \href
  {http://adsabs.harvard.edu/abs/2007MNRAS.374..787H} {374, 787}

\bibitem[\protect\citeauthoryear{{Heckman}, {Alexandroff}, {Borthakur},
  {Overzier}  \& {Leitherer}}{{Heckman} et~al.}{2015}]{Heckmea15}
{Heckman} T.~M.,  {Alexandroff} R.~M.,  {Borthakur} S.,  {Overzier} R.,
  {Leitherer} C.,  2015, \mn@doi [\apj] {10.1088/0004-637X/809/2/147}, \href
  {https://ui.adsabs.harvard.edu/abs/2015ApJ...809..147H} {809, 147}

\bibitem[\protect\citeauthoryear{{Iyudin} et~al.,}{{Iyudin}
  et~al.}{1994}]{Iyudea94}
{Iyudin} A.~F.,  et~al., 1994, \aap, \href
  {https://ui.adsabs.harvard.edu/abs/1994A&A...284L...1I} {284, L1}

\bibitem[\protect\citeauthoryear{{Jaskot}, {Strickland}, {Oey}, {Chu}  \&
  {Garc{\'{\i}}a-Segura}}{{Jaskot} et~al.}{2011}]{Jaskea11}
{Jaskot} A.~E.,  {Strickland} D.~K.,  {Oey} M.~S.,  {Chu} Y.-H.,
  {Garc{\'{\i}}a-Segura} G.,  2011, \mn@doi [\apj]
  {10.1088/0004-637X/729/1/28}, \href
  {http://adsabs.harvard.edu/abs/2011ApJ...729...28J} {729, 28}

\bibitem[\protect\citeauthoryear{{Knie}, {Korschinek}, {Faestermann}, {Dorfi},
  {Rugel}  \& {Wallner}}{{Knie} et~al.}{2004}]{Kniea04}
{Knie} K.,  {Korschinek} G.,  {Faestermann} T.,  {Dorfi} E.~A.,  {Rugel} G.,
  {Wallner} A.,  2004, \mn@doi [\prl] {10.1103/PhysRevLett.93.171103}, \href
  {https://ui.adsabs.harvard.edu/abs/2004PhRvL..93q1103K} {93, 171103}

\bibitem[\protect\citeauthoryear{{Kobayashi}, {Springel}  \&
  {White}}{{Kobayashi} et~al.}{2007}]{KSW07}
{Kobayashi} C.,  {Springel} V.,   {White} S. D.~M.,  2007, \mn@doi [\mnras]
  {10.1111/j.1365-2966.2007.11555.x}, \href
  {https://ui.adsabs.harvard.edu/abs/2007MNRAS.376.1465K} {376, 1465}

\bibitem[\protect\citeauthoryear{{Krause} \& {Diehl}}{{Krause} \&
  {Diehl}}{2014}]{KD14}
{Krause} M.~G.~H.,  {Diehl} R.,  2014, \mn@doi [\apjl]
  {10.1088/2041-8205/794/2/L21}, \href
  {http://adsabs.harvard.edu/abs/2014ApJ...794L..21K} {794, L21}

\bibitem[\protect\citeauthoryear{{Krause}, {Fierlinger}, {Diehl}, {Burkert},
  {Voss}  \& {Ziegler}}{{Krause} et~al.}{2013}]{Krausea13a}
{Krause} M.,  {Fierlinger} K.,  {Diehl} R.,  {Burkert} A.,  {Voss} R.,
  {Ziegler} U.,  2013, \mn@doi [\aap] {10.1051/0004-6361/201220060}, \href
  {http://adsabs.harvard.edu/abs/2013A\%26A...550A..49K} {550, A49}

\bibitem[\protect\citeauthoryear{{Krause}, {Diehl}, {B{\"o}hringer}, {Freyberg}
   \& {Lubos}}{{Krause} et~al.}{2014}]{Krausea14a}
{Krause} M.,  {Diehl} R.,  {B{\"o}hringer} H.,  {Freyberg} M.,   {Lubos} D.,
  2014, \mn@doi [\aap] {10.1051/0004-6361/201423871}, \href
  {http://adsabs.harvard.edu/abs/2014A%26A...566A..94K} {566, A94}

\bibitem[\protect\citeauthoryear{{Krause} et~al.,}{{Krause}
  et~al.}{2015}]{Krausea15a}
{Krause} M.~G.~H.,  et~al., 2015, \mn@doi [\aap] {10.1051/0004-6361/201525847},
  \href {http://adsabs.harvard.edu/abs/2015A%26A...578A.113K} {578, A113}

\bibitem[\protect\citeauthoryear{{Krause} et~al.,}{{Krause}
  et~al.}{2018}]{Krausea18b}
{Krause} M.~G.~H.,  et~al., 2018, \mn@doi [\aap] {10.1051/0004-6361/201732416},
  \href {http://adsabs.harvard.edu/abs/2018A%26A...619A.120K} {619, A120}

\bibitem[\protect\citeauthoryear{{Krause}, {Hardcastle}  \& {Shabala}}{{Krause}
  et~al.}{2019}]{Krausea19b}
{Krause} M. G.~H.,  {Hardcastle} M.~J.,   {Shabala} S.~S.,  2019, \mn@doi
  [\aap] {10.1051/0004-6361/201935762}, \href
  {https://ui.adsabs.harvard.edu/abs/2019A&A...627A.113K} {627, A113}

\bibitem[\protect\citeauthoryear{{Kretschmer}, {Diehl}, {Krause}, {Burkert},
  {Fierlinger}, {Gerhard}, {Greiner}  \& {Wang}}{{Kretschmer}
  et~al.}{2013}]{Kretschea13}
{Kretschmer} K.,  {Diehl} R.,  {Krause} M.,  {Burkert} A.,  {Fierlinger} K.,
  {Gerhard} O.,  {Greiner} J.,   {Wang} W.,  2013, \mn@doi [\aap]
  {10.1051/0004-6361/201322563}, \href
  {http://adsabs.harvard.edu/abs/2013A%26A...559A..99K} {559, A99}

\bibitem[\protect\citeauthoryear{{Krumholz}, {Burkhart}, {Forbes}  \&
  {Crocker}}{{Krumholz} et~al.}{2018}]{Krumhea18}
{Krumholz} M.~R.,  {Burkhart} B.,  {Forbes} J.~C.,   {Crocker} R.~M.,  2018,
  \mn@doi [\mnras] {10.1093/mnras/sty852}, \href
  {https://ui.adsabs.harvard.edu/abs/2018MNRAS.477.2716K} {477, 2716}

\bibitem[\protect\citeauthoryear{{Lichtenberg}, {Golabek}, {Burn}, {Meyer},
  {Alibert}, {Gerya}  \& {Mordasini}}{{Lichtenberg} et~al.}{2019}]{Lichtea19}
{Lichtenberg} T.,  {Golabek} G.~J.,  {Burn} R.,  {Meyer} M.~R.,  {Alibert} Y.,
  {Gerya} T.~V.,   {Mordasini} C.,  2019, \mn@doi [Nature Astronomy]
  {10.1038/s41550-018-0688-5}, \href
  {https://ui.adsabs.harvard.edu/abs/2019NatAs...3..307L} {3, 307}

\bibitem[\protect\citeauthoryear{{Limongi} \& {Chieffi}}{{Limongi} \&
  {Chieffi}}{2006}]{LC06}
{Limongi} M.,  {Chieffi} A.,  2006, \mn@doi [\apj] {10.1086/505164}, \href
  {https://ui.adsabs.harvard.edu/abs/2006ApJ...647..483L} {647, 483}

\bibitem[\protect\citeauthoryear{{Lugaro}, {Ott}  \& {Kereszturi}}{{Lugaro}
  et~al.}{2018}]{Lugarea18}
{Lugaro} M.,  {Ott} U.,   {Kereszturi} {\'A}.,  2018, \mn@doi [Progress in
  Particle and Nuclear Physics] {10.1016/j.ppnp.2018.05.002}, \href
  {https://ui.adsabs.harvard.edu/abs/2018PrPNP.102....1L} {102, 1}

\bibitem[\protect\citeauthoryear{{MacPherson}, {Bullock}, {Janney}, {Kita},
  {Ushikubo}, {Davis}, {Wadhwa}  \& {Krot}}{{MacPherson}
  et~al.}{2010}]{MacPhea10}
{MacPherson} G.~J.,  {Bullock} E.~S.,  {Janney} P.~E.,  {Kita} N.~T.,
  {Ushikubo} T.,  {Davis} A.~M.,  {Wadhwa} M.,   {Krot} A.~N.,  2010, \mn@doi
  [\apjl] {10.1088/2041-8205/711/2/L117}, \href
  {https://ui.adsabs.harvard.edu/abs/2010ApJ...711L.117M} {711, L117}

\bibitem[\protect\citeauthoryear{{Mackey}, {Gvaramadze}, {Mohamed}  \&
  {Langer}}{{Mackey} et~al.}{2015}]{MackeyJea15}
{Mackey} J.,  {Gvaramadze} V.~V.,  {Mohamed} S.,   {Langer} N.,  2015, \mn@doi
  [\aap] {10.1051/0004-6361/201424716}, \href
  {http://cdsads.u-strasbg.fr/abs/2015A%26A...573A..10M} {573, A10}

\bibitem[\protect\citeauthoryear{{Maiolino} \& {Mannucci}}{{Maiolino} \&
  {Mannucci}}{2019}]{MM19}
{Maiolino} R.,  {Mannucci} F.,  2019, \mn@doi [\aapr]
  {10.1007/s00159-018-0112-2}, \href
  {https://ui.adsabs.harvard.edu/abs/2019A&ARv..27....3M} {27, 3}

\bibitem[\protect\citeauthoryear{{Marshall}, {Shabala}, {Krause}, {Pimbblet},
  {Croton}  \& {Owers}}{{Marshall} et~al.}{2018}]{Marshea18}
{Marshall} M.~A.,  {Shabala} S.~S.,  {Krause} M. G.~H.,  {Pimbblet} K.~A.,
  {Croton} D.~J.,   {Owers} M.~S.,  2018, \mn@doi [\mnras]
  {10.1093/mnras/stx2996}, \href
  {https://ui.adsabs.harvard.edu/abs/2018MNRAS.474.3615M} {474, 3615}

\bibitem[\protect\citeauthoryear{{Nomoto}, {Kobayashi}  \& {Tominaga}}{{Nomoto}
  et~al.}{2013}]{NKT13}
{Nomoto} K.,  {Kobayashi} C.,   {Tominaga} N.,  2013, \mn@doi [\araa]
  {10.1146/annurev-astro-082812-140956}, \href
  {https://ui.adsabs.harvard.edu/abs/2013ARA&A..51..457N} {51, 457}

\bibitem[\protect\citeauthoryear{{Oey} \& {Garc{\'{\i}}a-Segura}}{{Oey} \&
  {Garc{\'{\i}}a-Segura}}{2004}]{OG04}
{Oey} M.~S.,  {Garc{\'{\i}}a-Segura} G.,  2004, \mn@doi [\apj]
  {10.1086/421483}, \href {http://adsabs.harvard.edu/abs/2004ApJ...613..302O}
  {613, 302}

\bibitem[\protect\citeauthoryear{{Orr} et~al.,}{{Orr} et~al.}{2020}]{Orrea20}
{Orr} M.~E.,  et~al., 2020, \mn@doi [\mnras] {10.1093/mnras/staa1619}, \href
  {https://ui.adsabs.harvard.edu/abs/2020MNRAS.496.1620O} {496, 1620}

\bibitem[\protect\citeauthoryear{{Pettitt}, {Ragan}  \& {Smith}}{{Pettitt}
  et~al.}{2020}]{PRS20}
{Pettitt} A.~R.,  {Ragan} S.~E.,   {Smith} M.~C.,  2020, \mn@doi [\mnras]
  {10.1093/mnras/stz3155}, \href
  {https://ui.adsabs.harvard.edu/abs/2020MNRAS.491.2162P} {491, 2162}

\bibitem[\protect\citeauthoryear{{Pleintinger}, {Siegert}, {Diehl}, {Fujimoto},
  {Greiner}, {Krause}  \& {Krumholz}}{{Pleintinger}
  et~al.}{2019}]{Pleintea19aph}
{Pleintinger} M. M.~M.,  {Siegert} T.,  {Diehl} R.,  {Fujimoto} Y.,  {Greiner}
  J.,  {Krause} M. G.~H.,   {Krumholz} M.~R.,  2019, arXiv e-prints, \href
  {https://ui.adsabs.harvard.edu/abs/2019arXiv191006112P} {p. arXiv:1910.06112}

\bibitem[\protect\citeauthoryear{{Pl{\"u}schke} et~al.,}{{Pl{\"u}schke}
  et~al.}{2001}]{Plea01}
{Pl{\"u}schke} S.,  et~al., 2001, in {Gimenez} A.,  {Reglero} V.,   {Winkler}
  C.,  eds,  ESA Special Publication Vol. 459, Exploring the Gamma-Ray
  Universe. pp 55--58 (\mn@eprint {arXiv} {astro-ph/0104047})

\bibitem[\protect\citeauthoryear{{Portegies Zwart}}{{Portegies
  Zwart}}{2019}]{PZ19}
{Portegies Zwart} S.,  2019, \mn@doi [\aap] {10.1051/0004-6361/201833974},
  \href {https://ui.adsabs.harvard.edu/abs/2019A&A...622A..69P} {622, A69}

\bibitem[\protect\citeauthoryear{{Prantzos} \& {Diehl}}{{Prantzos} \&
  {Diehl}}{1996}]{PD96}
{Prantzos} N.,  {Diehl} R.,  1996, \mn@doi [\physrep]
  {10.1016/0370-1573(95)00055-0}, \href
  {http://cdsads.u-strasbg.fr/abs/1996PhR...267....1P} {267, 1}

\bibitem[\protect\citeauthoryear{{Renaud} et~al.,}{{Renaud}
  et~al.}{2006}]{Renaudea06}
{Renaud} M.,  et~al., 2006, \mn@doi [\apjl] {10.1086/507300}, \href
  {https://ui.adsabs.harvard.edu/abs/2006ApJ...647L..41R} {647, L41}

\bibitem[\protect\citeauthoryear{{Renzini} \& {Andreon}}{{Renzini} \&
  {Andreon}}{2014}]{RA14}
{Renzini} A.,  {Andreon} S.,  2014, \mn@doi [\mnras] {10.1093/mnras/stu1689},
  \href {https://ui.adsabs.harvard.edu/abs/2014MNRAS.444.3581R} {444, 3581}

\bibitem[\protect\citeauthoryear{{Rodgers-Lee}, {Krause}, {Dale}  \&
  {Diehl}}{{Rodgers-Lee} et~al.}{2019}]{RodgLea19}
{Rodgers-Lee} D.,  {Krause} M.~G.~H.,  {Dale} J.,   {Diehl} R.,  2019, \mn@doi
  [\mnras] {10.1093/mnras/stz2708}, \href
  {https://ui.adsabs.harvard.edu/abs/2019MNRAS.490.1894R} {490, 1894}

\bibitem[\protect\citeauthoryear{{Rogers} \& {Pittard}}{{Rogers} \&
  {Pittard}}{2013}]{RP13}
{Rogers} H.,  {Pittard} J.~M.,  2013, \mn@doi [\mnras] {10.1093/mnras/stt255},
  \href {http://cdsads.u-strasbg.fr/abs/2013MNRAS.431.1337R} {431, 1337}

\bibitem[\protect\citeauthoryear{{Sasaki}, {Breitschwerdt}, {Baumgartner}  \&
  {Haberl}}{{Sasaki} et~al.}{2011}]{Sasea11}
{Sasaki} M.,  {Breitschwerdt} D.,  {Baumgartner} V.,   {Haberl} F.,  2011,
  \mn@doi [\aap] {10.1051/0004-6361/201015866}, \href
  {http://adsabs.harvard.edu/abs/2011A%26A...528A.136S} {528, A136+}

\bibitem[\protect\citeauthoryear{{Schulreich}, {Breitschwerdt}, {Feige}  \&
  {Dettbarn}}{{Schulreich} et~al.}{2018}]{Schulrea18a}
{Schulreich} M.,  {Breitschwerdt} D.,  {Feige} J.,   {Dettbarn} C.,  2018,
  \mn@doi [Galaxies] {10.3390/galaxies6010026}, \href
  {http://cdsads.u-strasbg.fr/abs/2018Galax...6...26S} {6, 26}

\bibitem[\protect\citeauthoryear{{Sellwood}, {Trick}, {Carlberg}, {Coronado}
  \& {Rix}}{{Sellwood} et~al.}{2019}]{Sellwea19}
{Sellwood} J.~A.,  {Trick} W.~H.,  {Carlberg} R.~G.,  {Coronado} J.,   {Rix}
  H.-W.,  2019, \mn@doi [\mnras] {10.1093/mnras/stz140}, \href
  {https://ui.adsabs.harvard.edu/abs/2019MNRAS.484.3154S} {484, 3154}

\bibitem[\protect\citeauthoryear{{Simionescu} et~al.,}{{Simionescu}
  et~al.}{2019}]{Simionea19}
{Simionescu} A.,  et~al., 2019, \mn@doi [\mnras] {10.1093/mnras/sty3220}, \href
  {https://ui.adsabs.harvard.edu/abs/2019MNRAS.483.1701S} {483, 1701}

\bibitem[\protect\citeauthoryear{{Sutherland} \& {Dopita}}{{Sutherland} \&
  {Dopita}}{1993}]{SD93}
{Sutherland} R.~S.,  {Dopita} M.~A.,  1993, \apjs, 88, 253

\bibitem[\protect\citeauthoryear{{Taylor} \& {Kobayashi}}{{Taylor} \&
  {Kobayashi}}{2015}]{TK15}
{Taylor} P.,  {Kobayashi} C.,  2015, \mn@doi [\mnras] {10.1093/mnrasl/slv087},
  \href {https://ui.adsabs.harvard.edu/abs/2015MNRAS.452L..59T} {452, L59}

\bibitem[\protect\citeauthoryear{{Toomre}}{{Toomre}}{1964}]{Toomre64}
{Toomre} A.,  1964, \mn@doi [\apj] {10.1086/147861}, \href
  {https://ui.adsabs.harvard.edu/abs/1964ApJ...139.1217T} {139, 1217}

\bibitem[\protect\citeauthoryear{{Tully}, {Courtois}, {Hoffman}  \&
  {Pomar{\`e}de}}{{Tully} et~al.}{2014}]{Tulea14}
{Tully} R.~B.,  {Courtois} H.,  {Hoffman} Y.,   {Pomar{\`e}de} D.,  2014,
  \mn@doi [\nat] {10.1038/nature13674}, \href
  {https://ui.adsabs.harvard.edu/abs/2014Natur.513...71T} {513, 71}

\bibitem[\protect\citeauthoryear{{Wallner} et~al.,}{{Wallner}
  et~al.}{2015}]{Wallnea15}
{Wallner} A.,  et~al., 2015, \mn@doi [Nature Communications]
  {10.1038/ncomms6956}, \href
  {https://ui.adsabs.harvard.edu/abs/2015NatCo...6.5956W} {6, 5956}

\bibitem[\protect\citeauthoryear{{Wang} et~al.,}{{Wang}
  et~al.}{2007}]{WangWea07}
{Wang} W.,  et~al., 2007, \mn@doi [\aap] {10.1051/0004-6361:20066982}, \href
  {https://ui.adsabs.harvard.edu/abs/2007A&A...469.1005W} {469, 1005}

\bibitem[\protect\citeauthoryear{{Wang} et~al.,}{{Wang}
  et~al.}{2020}]{WangWea20}
{Wang} W.,  et~al., 2020, \mn@doi [\apj] {10.3847/1538-4357/ab6336}, \href
  {https://ui.adsabs.harvard.edu/abs/2020ApJ...889..169W} {889, 169}

\bibitem[\protect\citeauthoryear{{Weaver}, {McCray}, {Castor}, {Shapiro}  \&
  {Moore}}{{Weaver} et~al.}{1977}]{Weavea77}
{Weaver} R.,  {McCray} R.,  {Castor} J.,  {Shapiro} P.,   {Moore} R.,  1977,
  \mn@doi [\apj] {10.1086/155692}, \href
  {http://adsabs.harvard.edu/abs/1977ApJ...218..377W} {218, 377}

\bibitem[\protect\citeauthoryear{{Woosley} \& {Heger}}{{Woosley} \&
  {Heger}}{2007}]{WH07}
{Woosley} S.~E.,  {Heger} A.,  2007, \mn@doi [\physrep]
  {10.1016/j.physrep.2007.02.009}, \href
  {https://ui.adsabs.harvard.edu/abs/2007PhR...442..269W} {442, 269}

\bibitem[\protect\citeauthoryear{{Zinnecker} \& {Yorke}}{{Zinnecker} \&
  {Yorke}}{2007}]{ZY07}
{Zinnecker} H.,  {Yorke} H.~W.,  2007, \mn@doi [\araa]
  {10.1146/annurev.astro.44.051905.092549}, \href
  {http://adsabs.harvard.edu/abs/2007ARA%26A..45..481Z} {45, 481}

\bibitem[\protect\citeauthoryear{{de Avillez} \& {Breitschwerdt}}{{de Avillez}
  \& {Breitschwerdt}}{2004}]{dAB04}
{de Avillez} M.~A.,  {Breitschwerdt} D.,  2004, \mn@doi [\aap]
  {10.1051/0004-6361:200400025}, \href
  {http://adsabs.harvard.edu/abs/2004A%26A...425..899D} {425, 899}

\bibitem[\protect\citeauthoryear{{von Glasow}, {Krause}, {Sommer-Larsen}  \&
  {Burkert}}{{von Glasow} et~al.}{2013}]{vGlea13}
{von Glasow} W.,  {Krause} M.~G.~H.,  {Sommer-Larsen} J.,   {Burkert} A.,
  2013, \mn@doi [\mnras] {10.1093/mnras/stt1060}, \href
  {http://adsabs.harvard.edu/abs/2013MNRAS.434.1151V} {434, 1151}

\makeatother
\end{thebibliography}




\bsp	
\label{lastpage}
\end{document}